\documentclass[sigconf]{acmart}
\newcommand{\remove}[1]{}
\usepackage{graphicx}
\usepackage{xcolor}
\usepackage{lipsum}
\usepackage{subfig}
\usepackage{amsmath}
\usepackage{mathtools}
\usepackage{mathrsfs}
\usepackage{tcolorbox}
\usepackage{enumitem}
\usepackage[noend]{algorithmic} 
\usepackage{algorithm}
\usepackage{flushend}
\usepackage{balance}
\usepackage{pifont}
\usepackage{tcolorbox}

\AtBeginDocument{%
  \providecommand\BibTeX{{%
    \normalfont B\kern-0.5em{\scshape i\kern-0.25em b}\kern-0.8em\TeX}}}

\usepackage{listings}
\definecolor{eclipseStrings}{RGB}{42,0.0,255}
\definecolor{eclipseKeywords}{RGB}{127,0,85}
\definecolor{codegray}{rgb}{0.5,0.5,0.5}
\colorlet{numb}{magenta!60!black}

\lstdefinelanguage{json}{
    basicstyle=\ttfamily\scriptsize,
    commentstyle=\color{black}, 
    stringstyle=\color{eclipseKeywords}, 
    numbers=left,
    numberstyle=\scriptsize,
    stepnumber=1,
    numbersep=8pt,
    showstringspaces=false,
    breaklines=true,
    string=[s]{"}{"},
    comment=[l]{:\ "},
    morecomment=[l]{:"},
    literate=
        *{0}{{{\color{numb}0}}}{1}
         {1}{{{\color{numb}1}}}{1}
         {2}{{{\color{numb}2}}}{1}
         {3}{{{\color{numb}3}}}{1}
         {4}{{{\color{numb}4}}}{1}
         {5}{{{\color{numb}5}}}{1}
         {6}{{{\color{numb}6}}}{1}
         {7}{{{\color{numb}7}}}{1}
         {8}{{{\color{numb}8}}}{1}
         {9}{{{\color{numb}9}}}{1}
}

\renewcommand\footnotetextcopyrightpermission[1]{} 
\copyrightyear{2022} 
\acmYear{2022} 

\setcopyright{none} 

\acmConference[SACMAT '22]{Proceedings of the 27th ACM Symposium on Access Control Models and Technologies}{June 8--10, 2022}{New York, NY, USA}
\acmBooktitle{Proceedings of the 27th ACM Symposium on Access Control Models and Technologies (SACMAT '22), June 8--10, 2022, New York, NY, USA}
\acmDOI{10.1145/3532105.3535014}
\acmISBN{978-1-4503-9357-7/22/06}

\usepackage{etoolbox}
\makeatletter
\patchcmd{\maketitle}{\@copyrightpermission}{
   \begin{minipage}{0.3\columnwidth}
     \href{http://creativecommons.org/licenses/by/4.0/}{\includegraphics[width=0.90\textwidth]{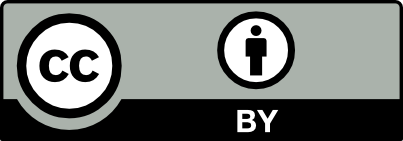}}
   \end{minipage}\hfill
   \begin{minipage}{0.7\columnwidth}
     \href{http://creativecommons.org/licenses/by/4.0/}{This work is licensed under a Creative Commons Attribution International 4.0 License.}
   \end{minipage}
  
   \vspace{5pt}
}{}{}

\makeatother

\begin{document}
\fancyhead{}
\setlength{\parskip}{0mm}
\title[]{ A Capability-based Distributed Authorization System to Enforce Context-aware Permission Sequences  }




\author{Adrian Shuai Li}
\authornote{Work completed while Adrian Shuai Li was at the University of Calgary.}
\affiliation{%
  \institution{Purdue University}
  \city{West Lafayette}
  \state{IN}
  \country{USA}}
\email{li3944@purdue.edu}
\author{Reihaneh Safavi-Naini}
\affiliation{%
  \institution{University of Calgary}
  \city{Calgary}
  \state{AB}
  \country{Canada}}
\email{rei@ucalgary.ca}

\author{Philip W. L. Fong}
\affiliation{%
  \institution{University of Calgary}
  \city{Calgary}
  \state{AB}
  \country{Canada}}
\email{pwlfong@ucalgary.ca}

\begin{abstract}
    Controlled sharing is fundamental to distributed systems. We consider a capability-based distributed authorization system where a client receives capabilities (access tokens) from an authorization server to access the resources of resource servers. Capability-based authorization systems have been widely used on the Web, in mobile applications and other distributed systems.
    
    A common requirement of such systems is that the user uses tokens of multiple servers in a particular order. A related requirement is the token may be used if certain environmental conditions hold.	We introduce a secure capability-based system that supports ``permission sequence''  and ``context''. This allows a finite sequence of permissions to be enforced, each with their own specific context. We prove the safety property of this system for these conditions and integrate the system into OAuth 2.0 with proof-of-possession tokens. We evaluate our implementation and compare it with plain OAuth with respect to the average time for obtaining an authorization token and acquiring access to the resource.

\end{abstract}




\maketitle

\section{Introduction}
\label{intro}

Securing access to protected resources is one of the most fundamental challenges in today's electronic world. Access control systems ensure that only access requests that match the stated security policies of the system are granted. Early access control systems considered centralized systems where a single computer system mediated and controlled access to the resources. Such centralized systems become a bottleneck today as the number of systems and services being protected is rapidly increasing. The poor scaling of centralized authorization systems stems from the manner in which every access request to the protected resources has to go back to the centralized system for access evaluation. 

Distributed authorization systems overcome the requirement of contacting the authorization system per request and thus provide a better approach for protecting large-scale systems. Distributed authorization in general, has two phases. In the first phase, a trusted {\em Authorization Server (\textbf{AS})} issues a capability that can be thought of as a token to a client after authenticating and verifying the request against the system policy, and in the second phase, the client hands over the capability to a trusted {\em Resource Server (\textbf{RS})}  that meditates the access to the protected resource. The capability must be unforgeable to ensure secure access.

A capability in its basic form is a random character string without any identifying information and so can be passed on to others, allowing unauthorized access. The capability can also be stolen and used without the RS noticing it. Gong \cite{gong1989secure} proposed an Identity-Based Capability System (ICAP)  that ties the capability to the identity of the user. The RS will verify the identity of the presenter  which results in the capability to fail the validity check. More recent capability systems allow actions, permissions, and auxiliary authorization information  to be encoded in the capability and enables RS to efficiently enforce
complex authorization decision using the capability itself, and 
without interacting with the AS.  OAuth 2.0 \cite{hardt2012oauth} is a prominent example of such a capability system. OAuth uses ``tokens'' to provide authentication and authorization in distributed settings including mobile applications \cite{chen2014oauth,shehab2014towards} and web services \cite{fett2016comprehensive, sun2012devil}. In this paper, we use the notions of capability and token interchangeably. 

Besides the progress we have seen for capability-based systems in recent years, existing systems do not offer control over orderings of permissions or limit the number of permission use. Thus delegated permissions can be exercised with arbitrary order and for unlimited number of times within a given period. In many applications, this  presents a severe security problem.  For example, in decentralized business and financial systems,  payment workflows require approvals of different authorities  in a particular order. Escrowing funds and seeking the correct  approval order 
ensures that the payment will be received by the intended recipient under the correct sets of checks. As a second example, in the Industrial Control Systems (ICS), the ordering of permissions to operate electronic equipment must conform to the workflow sequence \cite{tandon2018hcap}.  We note that not carefully controlled access to critical assets, even in a short window of time, can undermine security by allowing the attackers to deploy more sophisticated
attacks \cite{Intel, Target}. 

We propose an efficient system that removes the above weaknesses using a unified problem that can be stated as enforcing a finite sequence of permissions. We design capabilities that include the finite permission sequence and demonstrate how it can be updated in each step of showing the capability to an RS. The RS only needs to maintain an internal counter that will be advanced each time it receives a capability. We formally establish the security guarantees of the proposed system in the form of a safety property \cite{lamport1977proving} -- a property stipulating that no violation of the sequence constraint will occur. We further show that any protocol event that causes a state change in the distributed system can be mapped to an event that causes a state change in the centralized reference monitor.

In addition, our capability-based system includes the ``context''  of access. The term context refers to  any external conditions in the policies \cite{nissenbaum2009privacy}. Context information, including the state of the environment (e.g., ``turn on the home camera when the user is not home'') adds significant expressibility to the access control system. It enables a more refined expression of situations where access can be granted. Our technique leverages lessons from  Schuster et al. \cite{schuster2018situational}, which implemented a context server called \textit{Environmental Situational Oracle (ESO)}. An ESO encapsulates the implementation of how a situation is sensed, inferred, or
actuated \cite{schuster2018situational}. The integration of ESOs with our capability-based system enables control over any permission sequences with context. This allows a sequence of permission to be enforced, each with their specific context. Our security proof is still valid with the addition of context confinement.

We implement our capability system as an extension of OAuth 2.0, which shows our proposed system can strengthen OAuth to enforce context-aware permission sequences in distributed financial systems. We evaluate our system and compare it with OAuth 2.0 in terms of the average response time for two requests -- the request for obtaining authorization from AS and the request for accessing resources from RS. 
Our experiment shows that the overhead necessary for authorization requests is at most 5\%  when using the ECDSA token signature algorithm. However,  the overhead is a bit higher in the resource requests but remains a small constant. 

This paper is organized as follows. In Section \ref{overview}, we provide a high-level overview of our system. We present the design for enabling permission sequence in Section \ref{permission} and then show the extension that supports context-awareness in Section \ref{Context}. We discuss the limitations of our approach in Section \ref{limit}. We then describe the implementation details and the use case in Section \ref{imple}. We discuss the results of our experiments in Section \ref{eva}. We present related work in Section \ref{work} and conclude in Section \ref{Con}.

\section{System description}\label{overview}
\paragraph{\textbf{Protocol Participants}}

The proposed capability system considers the following entities \ding{202} \textit{Resource Owner (RO)} which is the entity owning an account in one or several resource servers. \ding{203} \textit{Resource Server (RS)} which is a server hosting RO's resources and providing services. We consider multiple resource servers where each RS hosts different resources.
\ding{204} \textit{Clients} which are the users  who wish to access the resource and/or  invoke permissions hosted on an RS by using software applications. \ding{205} \textit{Authorization Server (AS)} which is a centralized entity that provides authentication and authorization for the client access request. The AS issues capabilities to the client, who presents them to an RS as the proof of authorization. \ding{206} \textit{Environmental Situation Oracle (ESO)} which is a situation tracker that encapsulates the tracking of one environmental condition by processing raw data from the user devices. We consider multiple ESOs where each one tracks a different environmental condition.  An RS contacts the ESOs to check the environmental conditions before releasing the resource to the client. The ESOs evaluate the situation and respond valid/invalid to the RS. $\ $Note that we focus on the design of secure access to ESOs in this work, we do not propose any underlying situational tracking methods.

\paragraph {\textbf{Trust Assumptions}}
Our assumptions are \ding{202} Each entity has a private key and a certified public key issued by the RO. \ding{203} Any trusted entity can get other entities' public keys if needed by some suitable and trustworthy means. \ding{204} The AS, the resource servers, and the ESOs are trusted entities. ESO adopts sufficient methods to evaluate the state of the context correctly. \ding{205} Clients are malicious. They attempt to gain unauthorized access to the resource servers. We do not consider credential lending---the registered clients do not share their private key with the others.

\begin{figure}[t]
	\setlength{\belowcaptionskip}{-4pt}
	\centering
	\includegraphics[width=0.9\columnwidth]{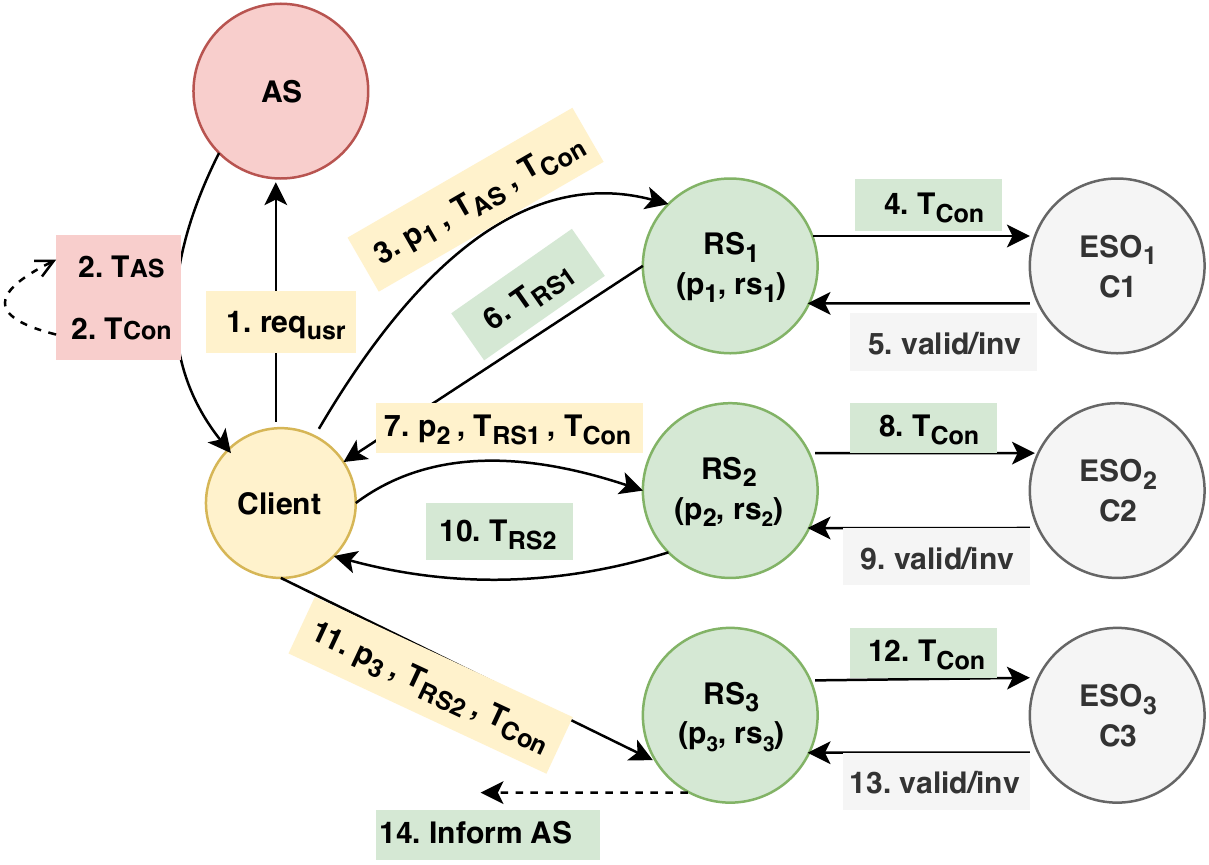}
	\caption[]{Generic flow diagram of the our system. The client must invoke permissions in the ordering of $\mathit{p_{1}, p_{2}, p_{3}}$. Each permission has its own context confinement. For the permission $p_{1}$, context $\mathit{C1}$ must be valid. For the permission $p_{2}$, context $\mathit{C2}$ must be valid. For the permission $p_{3}$, context $\mathit{C3}$ must be valid. After one run-through of the sequence, the AS is informed and it revokes $\mathit{T_{AS}}$.}
	\label{fig1}
\end{figure}
\paragraph{\textbf{Solution approach}}\label{sla}
First, we briefly introduce the OAuth 2.0 protocol. The client initiates the protocol by authenticating themselves and sending a request to the AS for accessing some resource that is controlled by an RS. The AS checks if the request is permitted by checking the local policies or asking RO's approval explicitly. If the check passes, the server then computes a capability  $\mathit{(t, tag_k(t))}$ and sends it to the client, where $\mathit{tag}$ is a digital signature signed by AS's private key, and $t$ is the data to be signed. The RS can verify $\mathit{tag}$ because they have the AS's public key.  An OAuth capability usually contains assertions related to authorization, for example, the granted permission, the expiration time of the capability, the issuer of the capability, the RS in which the capability should be used, and the client's identity. In this solution, permission sequence enforcement is not supported. Some simple contexts are supported, yet the conditions of the authorization can not be fully expressed.

To restrict the orderings in which the permissions are invoked, we include an assertion that specifies the permission sequence, and the associate resource servers in the capability. An example is shown in step 2 of Figure \ref{fig1}, where $T_{AS}$ includes those information for a  permission sequence $\mathit{p_{1}, p_{2}, p_{3}}$. The permission sequence can include different permissions (in the case of controlling orderings), or repeated permissions (in the case of limiting the number of access).  To enforce the sequence among multiple resource servers, we introduce a new capability issued by an RS whenever a permission has been invoked from that RS. The new capability includes a $\mathit{state}$ assertion, which asserts the next state (index) in the permission sequence. 

We give the following example to demonstrate the above process. Initially, the $\mathit{state}$ value in $T_{AS}$ is $0$, meaning the client should invoke the permission located at index 0 of the sequence. In step 3, the client requests the permission $p_1$ with $T_{AS}$. If the request is allowed (we will explain shortly on the process of verifying a request), $RS_1$ issues a capability $T_{RS_1}$ to the client. We advance the $\mathit{state}$ value by 1 in $T_{RS_1}$ to represent the next index in the permission sequence. Then, the client presents $T_{RS_1}$ to $RS_2$ in step 7 and gets another capability  $T_{RS_2}$.  The same interactions repeats until all permissions in the sequence are visited in a correct order.  

However, the above solution enables the client to replay these capabilities generated in one session and rewind the system state.  The fundamental problem is that resource servers do not track any access history. To resolve this issue, we let each RS keeps a local variable. Whenever an RS issues a new capability, that RS advances its local variable value to the new state. When the client presents any old capabilities an RS has already seen, the RS can detect the replay of capabilities by comparing the $\mathit{state}$ value in the capability with the local variable value in the RS.  We show the protocol details for enabling permission sequence constraint in Section \ref{permission}.

To verify the environmental context attached with each permission, we use an external context tracker called ESO. The access to the ESOs must be protected since it provides information related to the user and his surrounding environment. We let the AS control the access to the ESOs.  A natural approach is to present the capability $\mathit{T_{AS}}$ to the ESOs as proof of authorization. This approach has privacy concerns as the ESOs could gain  information about the permission sequence granted to the client. We resolve this issue by introducing an ESO capability called $T_{Con}$, which is the proof of authorization to access an ESO. A $T_{Con}$ is issued in pair with the $T_{AS}$ to the client. The policy in the $T_{AS}$ is not included in the $T_{Con}$.  Upon receiving both capabilities, the RS queries the ESOs using the ESO capability. The protocol details are discussed in Section \ref{Context}.

Final note on the efficient capability revocation in our system.  All  capabilities shown in Figure \ref{fig1} are either cryptographic tied to, or derived from $T_{AS}$ (see Section \ref{permission} and \ref{Context} for more details). Hence, revoking the $T_{AS}$ will invalidate all descendant capabilities.\\

\balance
\section{Permission Sequence}\label{permission}

In this section, we present the design for enabling permission sequence constraint using two underlying primitives, hash function and digital signatures. Then we formalize the protocol as a distributed state transition model. Further we prove that any protocol event that causes a state change in the distributed system can be mapped to a event that causes a corresponding state change in the centralized reference monitor (Section \ref{4.2}).\\

{\bf{\textit{Preliminaries}}}.
We use $\mathnormal{H}$ to denote a hash function. $\mathnormal{H}(m)$ is the digest of message $m$. 

 $\mathnormal{S}_A$ is a signing transformation for entity $A$ from the message set $\mathnormal{M}$ to the signature set $\mathnormal{S}$. The signature created by Entity $A$ for a message $m \in \mathnormal{M}$ is denoted as $\mathnormal{S}_{A}(m)$. $\mathnormal{V}_A$ is  a verification  transformation for $A$'s signatures from the set $\mathnormal{M} \times \mathnormal{S}$ to the set $\{\mathit{true,false}\}$. To verify $\mathnormal{S}_{A}(m)$ is  indeed created by $A$, entity $B$ computes $\mathnormal{V}_A(m,\mathnormal{S}_{A}(m) )$. The signature is accepted if the result is $\mathit{true}$, otherwise rejected. 

We use output $\perp$ to denote any failure response.  We use ``,'' for concatenation. 

\subsection{Protocol Description}\label{protocoldes1}

The protocol session starts when a client initiates a request to the AS until to the point when the capability is revoked or expired.  

{\bf \textit{The structures of capabilities}}. For each session, the capability $\mathcal{T}$ is a set which is composed of two elements: $\mathit{T_{AS}}$ - a master capability issued by the AS; $\mathit{T_{RS}}$ - a set of capabilities issued by resource servers. We call $\mathit{T_{RS}}$ the state capability set. Below we present the structure of each capability. 

\noindent\underline{$\mathit{T_{AS}}$}
\begin{itemize}[label={--}]
	\item $\mathit{T_{AS}}$ has the form of $(t,\mathnormal{S}_{AS}(t)) $, \\where $\mathit{t=(\mathcal{P}, C_{id}, state=0,session_{id},exp)}$.	
	\item  $\mathcal{P}$ contains a finite sequence of permissions and  the RS identifiers for each named permission. For the example in Figure \ref{fig1}, 
	We could express $\mathcal{P}$ as 
	\begin{equation}
		\begin{split}
			[\mathit{(RS_1,p_{1}), (RS_2,p_{2})},\mathit{(RS_3,p_{3})}]
		\end{split}
	\end{equation}
	\item Among the other assertions, $T_{AS}$ includes the identity of the client ($C_{id}$), session id, the state value of 0 which indicates that the permission at index 0 should be invoked next and the expiration time after which the capability is not valid.
	\item $\mathnormal{S}_{AS}(t) $ is the signature created by the AS for message $t$. 
\end{itemize}
\noindent\underline{$\mathit{T_{RS}}$}
\begin{itemize}[label={--}]
	
	\item The elements in this set are capabilities issued from the resource servers. A capability issued by $RS_{id}$ is denoted as $T_{RS_{id}}$. 
	\item The elements in this set has the form of $(t,\mathnormal{S}_{RS_{id}}(t)) $, where $\mathit{t=(T_{AS},  state,cert_{RS_{id}},exp^{})}$.
	\item $T_{RS_{id}}$ contains the master capability $T_{AS}$, certificate of $RS_{id}$ assuming that the resource servers does not necessarily know the other resource servers in the system.
	\item The state value indicates the index of the next permission in $\mathcal{P}$.
	\item The signature ensures the data authenticity (the RS is really who they claimed to be), and integrity. Upon receipt of an state capability $\mathit{T_{RS_{id}}}$, the next RS must verify that the signature of AS in $T_{AS}$ is valid, and signature of $RS_{id}$ is valid  using the certified public key.
\end{itemize}

\paragraph{\textbf{RS Internal State}}
Each RS maintains a local variable $rs$ for each session. $rs$ is updated whenever a permission that it holds is invoked successfully. More specifically, $\mathit{rs \xleftarrow[]{} state +1}$, where $\mathit{state}$  value is from the capability presented to the RS. The initial value of $rs$ for each session is 0. 

\paragraph{\textbf{Obtaining Authorization from AS}}
A client authenticates to the AS before he/she sends any authorization request. If the client authenticates successfully and it is authorized to perform requested permissions based on policies, the AS issues a capability  $T_{AS}$ to the client.  $T_{AS}$ has the form of $(t,\mathnormal{S}_{AS}(t))$, where $t=(\mathit{\mathcal{P}, C_{id}, state=0,}$\\$session_{id},$ $exp)$. 

\paragraph{\textbf{Accessing Protected Resource from RS}}
Client attempts to exercise permission $p$ on $RS_m$ by presenting a capability $T$, where $\mathit{T}$ is either $T_{AS}$ or from the set $\mathit{T_{RS}}$. After verifying the identity of the client, the RS runs Algorithm \ref{alg1} to handle the request. The RS first checks if the client presents a capability with access, and whether the signature is valid (lines 1-3). In line 4, the RS checks the other important assertions in the capability, including \ding{202} the identity of the client matches the  $C_{id}$ in the capability \ding{203} the capability is not expired \ding{204} the capability is not revoked by the AS, possibly by checking the revocation list which contains all the revoked capabilities, \ding{205} The requested permission $p$ is associated with $RS_m$.

\begin{algorithm}[t]
	\caption{Procedure of Authorization at RS}
	\label{alg1}
	\begin{algorithmic}[1]
		\renewcommand{\algorithmicrequire}{\textbf{Input:}}
		\renewcommand{\algorithmicensure}{\textbf{Output:}}
		\REQUIRE A client access request $(\mathit{C_{id}, p, T)}$ to $RS_{m}$.
		\ENSURE  A new capability, or $\perp$.
		\IF {$\mathit{T} $ is null}
		\RETURN $\perp$ 
		\ENDIF
		\IF {$\mathit{T}$ is $\mathit{T_{AS}}$ \OR $\mathit{T \in T_{RS}}$ }
		\IF {$\mathit{T}$ is not issued for $C_{id}$ \OR $\mathit{T}$ is expired \OR $\mathit{T}$ is revoked \OR $\mathit{T}$ is not intended to use at the $RS_m$ }
		\RETURN  $\perp$ 
		\ELSIF {$\mathit{state \ge rs_{m}}$ \AND invoking $p$ at $RS_m$ is allowed by the current state of $\mathcal{P}$  }
		\STATE Invoke $p$ 
		\ELSE{}
		\RETURN $\perp$ 
		\ENDIF
		
		\IF {$p$ is the last permission in $\mathcal{P}$}
		\STATE inform the AS 
		\ELSE 
		\STATE $\mathit{rs_{m} \xleftarrow[]{} state+1}$ 
		\RETURN $\mathit{T_{RS_{m}}}$
		\ENDIF
		\ELSE 
		\RETURN $\perp$
		\ENDIF
	\end{algorithmic}
\end{algorithm}

Only when all the checks above are passed, the RS starts to consider if the client follows the restricted order of permissions. In line 6, if the value of $\mathit{state}$ in the capability is larger than or equal to the value of $rs_m$ and invoking permission $p$ on $RS_m$ is allowed by the current state of $\mathcal{P}$, $RS_m$ invokes the permission $p$. $RS_m$ then updates the $rs_m$ to $\mathit{state+1}$ and creates a new capability $T_{RS_m}$ (lines 13-14).  $T_{RS_m}$ has the form   of $(t,S_{RS_{m}}(t))$, where $\mathit{t=(T_{AS}, state=rs_m,cert_{RS_{m}},exp^{})}$. The $state$ value in $T_{RS_m}$ equals to the updated $rs_m$ value. By doing this, the latest state value is kept by $RS_m$ and carried by the capability to the next resource server. Finally, if $p$ is the last permission, the RS informs the AS about the complete run-through of the permissions in $\mathcal{P}$. No new capability is generated. Then AS revokes the master capability, all the state capabilities will be invalid automatically since they contain the master capability. All resource servers delete their internal variables for this session. 

\subsection{Security Analysis}\label{4.2}
We provide security analysis for the following attacks:

{\bf{\textit{Capability forgery and tampering.}}}  The attacker tries to forge and/or tamper capabilities to gain unauthorized access. To prevent this attack, we include a digital signature in every capability.  The private key is used to create signatures that prove ownership of controlling access (AS) and resources (RS). Any forged/tampered capability will be detected during signature validation. 

{\bf{\textit{Capability theft.}}} The attacker may steal the capability in transit if communication channels are not protected by TLS/DTLS. Nevertheless, he can not use these stolen credentials because each capability includes the identity of the possessor (client id). When the client presents the capability to the RS, he must authenticate to the RS by proving possession of the private key only known by the possessor of the capability (e.g., by signing a protocol message with a private key). Using any capability issued for others will be detected by the server.  We recommend using TLS for privacy reasons. The capability is signed, but not encrypted. Although the attacker can not use the stolen capability, he may infer the security policy from the capability. 

{\bf{\textit{Client Impersonation.}}} The attacker impersonates as client A to obtain authorization from the AS. This attack is prevented by public-key-based client authentication. 

{\bf{\textit{Replay attack }}} is the main security concern here. During one session, client receives a set of capabilities, including one master capability and multiple state capabilities. Is it possible for the client to reuse any old capabilities that an RS has already seen?  Replaying an old capability could lead to unauthorized access which violates the permission sequence. \textit{ We formalize our protocol in the following and prove that replay attacks are not possible. }

We model the sequence enforcement protocol as a state transition system. Each protocol state captures the states of the entire distributed system, including states of the AS, resource servers, and the client. We then list all protocol events that cause a state transition in the distributed system. 
Next we model the state transition for enforcing permission sequence in a centralized reference monitor.  Finally, with the effective state (security property), we show that the distributed system simulates the behavior of a centralized system in \textit{\bf{Theorem 1}}. We use $|\mathcal{P}|$ to denote the length of permission sequence $\mathcal{P}$.

Our state transition model abstracts away the following aspects of the protocol: (1) Tickets forging is not modelled as we assume that it is adequately prevented by signatures. (2) The model specifies the behavior of one session only since each session is independent from one another. (3) Each capability allows one permission to be exercised at a time. \\

{\bf{\textit{Definition 1 (Protocol States)}}}. A protocol state $\gamma$ is a 3-tuple $\mathit{(A, R, C)}$, where the three components are defined as follows.
\begin{itemize}[label={--}]
	\item The authorization server state $A$ is a binary value $b$ that represents whether the original capability $T_0$ has been issued from the AS.  
	\item The state of resource servers is represented by $R$, where $R$ is a vector of counter values. $R=\{rs_{i}\}_{i \in [n]}, [n]={0,..., n-1}$. 
	\item The client state $C$ is the set of capabilities that have been issued to the client throughout the protocol session. A capability $\mathit{T}$ is of the form $\mathit{T (\mathcal{P}, \mathit{state})}$, where $\mathcal{P}$ is a sequence of permissions. Each permission is specified with the identity of RS on which the permission should be invoked. The second element  $\mathit{state}$ indicates the index of the next permission to be invoked in $\mathcal{P}$.
\end{itemize}
Let $\Gamma$ be the set of all protocol states $\gamma$ of the above form. \\

{\bf{\textit{Initial state.}} }The protocol is intended to begin at the initial state $\mathit{\gamma_{0}=(A_{0}, R_{0}, C_{0})}$ where $A_{0}$ is the initial state of the authorizaton server ($A_{0}=false$),$\ $ $R_0$ is the initial state of the resource servers ($\mathit{\forall i \in [n]\ rs_{i}=0}$),$\ $ $C_{0}$ is the initial set of capabilities that have been issued to the client ($C_{0} = \varnothing$). 
\\

{\bf{\textit{State transition.}}} A transition identifier $\lambda$ identifies a protocol event that causes a change to the protocol state: 
\begin{equation}
	\mathit{\lambda::=} \mathit{issue()\ |\ request(p,T, RS_{m})\ |\ }
	\mathit{last\_request(p,T, RS_{m})}
\end{equation}

where $p$ is a permission, $T$ is a capability, $RS_{m}$ is the resource server client attempts to visit, $m \in [n]$. Let $\Lambda$ be the set of all transition identifiers. 

We specify a transition relation $ \cdot \xrightarrow[]{\cdot}\cdot \subseteq \Gamma \times\Lambda \times\Gamma$. The relation is specified in terms of \textbf{transition rules}, which identify the conditions under which $(\mathit{A, R, C) \xrightarrow[]{\lambda} (A^{'}, R^{'}, C^{'})}$, where $\mathit{A = b}$, $R=\{rs_{i}\}_{i \in [n]}$, $\mathit{A^{'} = b^{'}}$, $R^{'}=\{rs_{i}^{'}\}_{i \in [n]}$. By default $\mathit{A^{'} = A}$, $\mathit{R^{'} = R}$, $\mathit{C^{'} = C}$, unless the rules explicitly say otherwise.

\begin{tcolorbox}[colback=gray!5!white,colframe=gray!75!black]
\noindent{\bf{T-Iss}} The AS issues a capability to the client. \\
\indent \textbf{Precondition:} $\mathit{\lambda = issue()}$, $\mathit{b=false}$.\\
\indent \textbf{Effect:} (i) $b^{'}=true$. (ii) $C^{'}=C \cup \{\mathit{T_{0}\}}$, where $\mathit{T_{0}= T\{\mathcal{P},\ }\mathit{ state=0\}}$.\\

\noindent{\bf{T-Req}} The client requests to exercise a permission. \\
\indent \textbf{Precondition:} $\mathit{\lambda = request(p,T,RS_{m})}$, $\mathit{T \in C}$, $\mathit{T =}$$\mathit{T (\mathcal{P}, }$
$\mathit{\mathit{ state)}}$, $\mathit{state \geq rs_{m}}$, $\mathit{p =\mathcal{P}[state]}$, $\mathit{state <}$$\mathit{ |\mathcal{P}-1|}$, $\mathit{b=true}$. \\

\indent \textbf{Effect:} (i) $\mathit{rs^{'}_{m} = state}+1$, $\forall i\in [n] \setminus	\{m\}\ rs^{'}_{i}=rs_{i}$.  (ii) $\mathit{C^{'}= C\ \cup}$
\indent$\mathit{ \{\mathit{T_{new}\}}}$, where $\mathit{T_{new}=\ }$$\mathit{T}$	$(\mathcal{P},\mathit{rs^{'}_m)}$.\\

\noindent{\bf{T-Com}} The RS informs the AS after client  requesting the last permission in $\mathcal{P}$.   \\
\indent\indent \textbf{Precondition:} $\mathit{\lambda = last\_request(p,T,}$$\mathit{RS_{m})}$, $\mathit{T \in C}$, $\mathit{T =}$ $\mathit{T (\mathcal{P}, state)}$, $\mathit{state \geq rs_{m}, }$ $p =$ $\mathit{\mathcal{P}[state]}$, $\mathit{state =}$$\mathit{ |\mathcal{P}-1|}$, $\mathit{b=true}$.\\
\textbf{Effect:} (i) $\mathit{R^{'}= R_{0}}$. (ii) $\mathit{A^{'}=A_{0}}$. 
\end{tcolorbox}

{\bf{\textit{State invariants:}}} 

Inv-1: When $b$ is true, for any $\mathit{T \in C}$, and any resource server $RS_{i}$, one of the following two cases holds: (a) if $RS_{i}$ does not match with the identity of the RS who holds permission $p$, then $T$ can not be used at $RS_{i}$; (b) if $RS_{i}$ matches with the identity of the RS who holds permission $p$, the $\mathit{state}$ value in $\mathit{T}$ is always no less than the value of $rs_{i}$ in $RS_{i}$. 


Inv-2: 
When $b$ is false, $R^{'}= R_{0}$, all capabilities in the session are revoked.\\ 

{\bf{\textit{Proposition 1 (State Invariants)}}}. The initial state $\gamma$ satisfies conditions Inv-1 to Inv-2. In addition, if $\gamma$ satisfies conditions Inv-1 to Inv-2, and $\gamma \xrightarrow[]{\lambda} \gamma^{'}$, then $\gamma^{'}$ also satisfies those conditions.

The proof of Proposition 1 is provided in \cite[Chapter 4, Section 2]{li2020capability}. \\

{\bf{\textit{Security property.}}} Consider a protocol state $\gamma = (A, R, C)$, such that $\gamma$ satisfies the Inv-1 to Inv-2. The \textbf{effective state} of protocol state $\gamma$, denoted $\mathit{eff}(\gamma)$, is defined below: 

\begin{equation}
	\mathit{eff}(\gamma)= 
	\begin{cases}
		0  & \text{if } \mathit{b = false}\\
		\mathit{max(rs_{i})_{i \in [n]}} & \text{otherwise}\\
	\end{cases}
\end{equation}\\

{\bf{\textit{Definition 2 (Centralized System States)}}}. A centralized system state is the value of  counter $ctr$ which indicates the index of the current permission in the sequence $\mathcal{P}$. The initial state of the system is $ctr=0$.  Let $\Gamma^{'}$ be the set of all states. \\

{\bf{\textit{State transition.}}} A transition identifier $\lambda^{'}$ identifies a protocol event that causes a change to the protocol change: 
\[\mathit{\lambda^{'}::=  request^{'}(p)\ |\ last\_request^{'}(p)}\]
where $p$ is a permission. Let $\Lambda^{'}$ be the set of all transition identifiers. 

We specify a transition relation $ \cdot \xrightarrow[]{\cdot}\cdot \subseteq \Gamma^{'} \times\Lambda^{'} \times\Gamma^{'}$. The relation is specified in terms of \textbf{transition rules}, which identify the conditions under which $ctr \xrightarrow[]{\lambda^{'}} ctr^{'}$. By default $ctr^{'}=ctr$, unless the rules explicitly say otherwise. 

\begin{tcolorbox}[colback=gray!5!white,colframe=gray!75!black]
{\bf{T-CReq}} The client requests to exercise a permission. \\
\indent\indent \textbf{Precondition:} $\mathit{\lambda^{'} = request^{'}(p)}$ 

\indent\indent \textbf{Effect:}:  $\mathit{ctr^{'}=ctr+1}$\\

{\bf{T-CCom}} The client requests to exercise the last permission in $\mathcal{P}$. \\
\indent\indent \textbf{Precondition:} $\lambda^{'} = last\_request^{'}(p)$ and $ctr=$$|\mathcal{P}-1|$

\indent\indent \textbf{Effect:}:  $ctr^{'}=0$
\end{tcolorbox}

{\bf{\textit{Theorem 1 (Safety)}}}. Suppose $\gamma$ satisfies the state invariant from Inv-1 to Inv-2, and $\gamma \xrightarrow[]{\lambda} \gamma^{'}$. Then the following statements hold:
\begin{itemize}[label={--}]
	\item if $\mathit{\lambda = request (p, T, RS_{m})}$, 
	\begin{itemize}
		\item[]$\mathit{eff(\gamma) \xrightarrow[]{where\ \lambda^{'} = request^{'}(p) }eff(\gamma^{'})}.$
		\end{itemize}
	
	\item  if $\mathit{\ \lambda = last\_request (p, T, RS_{m})}$,
	\begin{itemize} 
		\item[]$\mathit{eff(\gamma) \xrightarrow[]{\lambda^{'} = last\_request^{'}(p) }eff(\gamma^{'})}$.
	\end{itemize}
	\item if $\mathit{\lambda}$ is not of the form $\mathit{request(\_,\_,\_)}$, or $\mathit{last\_request(\_,\_,\_)}$, 
	\begin{itemize}
		\item[]$\mathit{eff(\gamma)=eff(\gamma^{'})}$.
	\end{itemize} 
	
\end{itemize}

The theorem above states that any protocol event that causes a state change in the distributed system can be mapped to a event that causes state change in the centralized reference monitor. Proof of Theorem 1 is provided in \cite[Chapter 4, Section 2]{li2020capability}.

\section{Enabling Context-awareness}\label{Context}
\begin{figure}[t]
	\centering
	\includegraphics[width=\columnwidth]{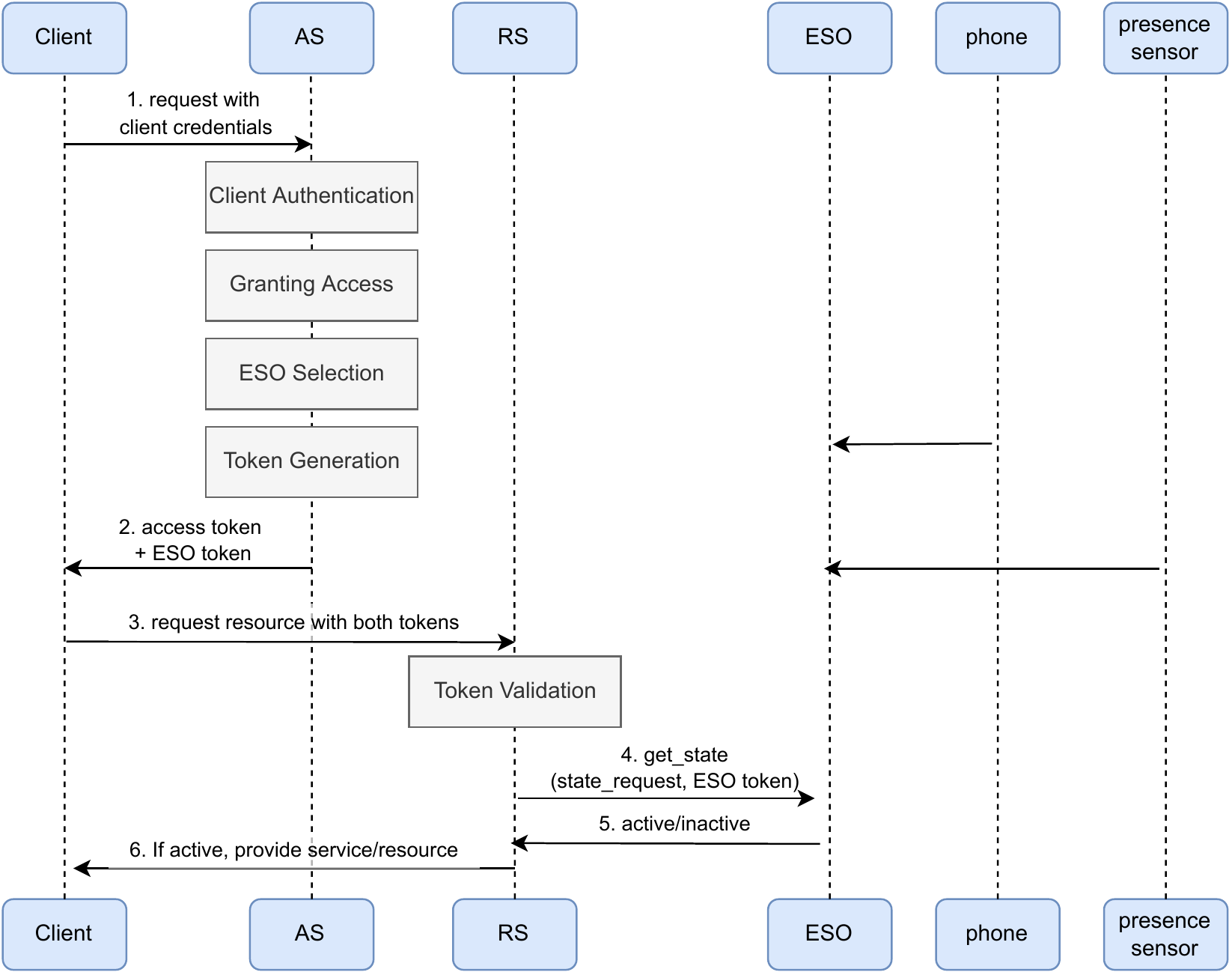}
	\caption{Authorization enforcement process with context.}
	\label{eso}
\end{figure}

In this section, we present the design for enabling context-awareness with the same preliminaries defined in Section \ref{permission}. Figure \ref{eso} shows the complete flow with context-awareness. The notions of capability and token are used interchangeably. 
 
\subsection{Protocol Description}

\paragraph{\textbf{The Structures of Capabilities}}

We introduce two new capabilities in the capability set $\mathcal{T}$ defined in Section \ref{protocoldes1}. The new capabilities are $\mathit{T_{Con}}$ - a capability issued by the AS for accessing the ESO and $\mathit{{T_C}}$ - a capability issued by the client for sending authorization request to the AS. We update the structure of $\mathit{T_{AS}}$ to enable the context feature. $\mathit{T_{RS}}$ keeps the same structure defined in Section \ref{protocoldes1}.

\noindent\underline{$\mathit{T_{AS}}$}
\begin{itemize}[label={--}]
	\item 
	$\mathit{T_{AS}}$ has the form of $(t,\mathcal{S}_{AS}(t)) $, \\where $\mathit{t=(\mathcal{P}, C_{id}, state=0 ,session_{id},exp)}$.
	\item $\mathcal{P}$ is an information structure specifies what are the authorized permissions, where to invoke these permissions, and what are the context associate with these permissions. Each permission can be specified as the tuple of three elements
	\begin{equation}\label{e1}
	(\mathit{RS_{id}, p, Context(C_{id},RS_{id},property)})
	\end{equation} 
	$\mathit{ Context(C_{id},RS_{id},property)}$ is a check/condition says that a ``property''  must be satisfied by the client with the identity $\mathit{C_{id}}$ and the RS  with the identity $\mathit{RS_{id}}$. The $\mathit{property}$ could be  as simple as when to invoke the permission or as complex as requiring certain location of the user. Note that one permission $p$ could possibly has several context constraints.  We give an example of $\mathcal{P}$  based on Figure \ref{fig1}. Suppose $\mathit{Context1}$, $\mathit{Context2}$, $\mathit{Context3}$ are three different context property. 
	We could express $\mathcal{P}$ as 

	\begin{equation}
	\begin{split}
	\mathcal{P}=[\mathit{(RS_1,p_{1},Context1 ), (RS_2,p_{2},Context2),}\\
	\mathit{(RS_3,p_{3},Context3 ),(RS_1,p_{1},Context1 )}]  
	\end{split}
	\end{equation}
	
\end{itemize}

\noindent\underline{$\mathit{T_{Con}}$}
\begin{itemize}[label={--}]
	\item  $\mathit{T_{Con}}$ has the form of $(t,\mathnormal{S}_{AS}(t)) $, \\where $\mathit{t=(\mathnormal{H}(T_{AS}), scope, exp )}$.
	\item  The $\mathnormal{H}(T_{AS})$ ties the $\mathit{T_{Con}}$ with the $\mathit{T_{AS}}$. The purpose  is that the client should not use the ESO capability without the master capability, or use it with a different master capability from another session. 
	\item  The $\mathit{scope}$ is specified as the tuple of four elements,
	\begin{equation}
	(\mathit{RS_{id}, ESO_{id}, p^{'}, Context})   
	\end{equation} 
	$\mathit{ESO_{id}}$ indicates the identity of the ESO which tracks the context. In the case of an URL is used for $\mathit{ESO_{id}}$ , there is no need to map from the ESO identity to its URL. $p^{'}$ indicates the authorized permission(s) on the ESO, which is usually ``read''. 
\end{itemize}

\noindent\underline{$\mathit{T_{C}}$}
\begin{itemize}
	\item $\mathit{T_{C}}$ has the form of $(t,\mathit{S}_{C}(t))$,\\ where $\mathit{t=}\mathit{((C_{id}), RS_{id}, requested\_scope)}$.
	\item  A request from the client is represented as a capability. The $\mathit{requested\_scope}$ indicates the permission(s) requested by the client. The $\mathit{requested\_scope}$ captures the rich semantics of authorization requests.
	\item Because of the signature, the AS is able to authenticate the client and check the integrity of the request.
\end{itemize}

\paragraph{\textbf{Obtaining Authorization from AS}}

Client $C_{id}$ sends an authorization request to the AS with a client claim capability $T_{C}$. This capability includes the request information and the client's signature for authentication. If the  client authenticates successfully  to the  AS  and  it is authorized  to perform requested permissions, the AS issues a master capability $T_{AS}$ and an ESO capability $T_{Con}$ to the client. 

The policies at AS should add ESO policies. One way of adding the ESO policy to the access control policy is to hard-code the ESO URL in policy rules, as specified in \cite{schuster2018situational}. ESO registration can be done by maintaining a database that stores the ESO description and its URL at the AS - we present more details of installing an ESO server in Section \ref{imple}.

\paragraph{\textbf{Accessing Protected Resource from RS}} 
Client attempts to exercise permission $p$ on $\mathit{RS_m}$ by presenting $(\mathit{ C_{id}, p, T,T_{Con})}$, where $\mathit{T}$ is either $T_{AS}$ or from the set $\mathit{T_{RS}}$. After verifying the identity of the client, the RS first checks if the client presents two unexpired capabilities with access.  Then, the RS verifies whether the ESO capability $T_{Con}$ is tied with the master capability $T_{AS}$. This can be easily done if $\mathit{T}$ is $T_{AS}$. If $\mathit{T}$ is from the set $\mathit{T_{RS}}$, the RS needs to find  $T_{AS}$ within the capability.

After all the above checks have passed, the RS 
examines if the client follows the restricted permission order as shown in lines 3 - 9 in Algorithm \ref{alg1}. The only modification is that before invoking the permission  $p$ in line 7, the RS queries the ESO server if a certain context has to be satisfied. Before the RS interacts with the ESO server, the identity of the RS must be verified.  RS calls $\mathit{get\_state()}$ method of the ESO  with  $\mathit{T_{Con}}$ to query the state of the context. When the ESO receives a request from the RS, it verifies that \ding {202} the signature of $\mathit{T_{Con}}$ is valid using AS's public key, \ding{203} the capability is not expired, \ding{204} accessing rights are allowed by $\mathit{T_{Con}}$. Finally, the ESO returns active/inactive to the RS, which indicates the result of the context evaluation. If the result is active, RS will invoke $p$ for the client. Otherwise, RS returns $\perp$. The rest of the steps follows lines 10 - 14 in Algorithm \ref{alg1}.

\subsection{Security Analysis }\label{5.2}
The context-aware permission sequence protocol prevents attacks mentioned in section \ref{4.2}. Any violation of the permission sequence by replaying the old token is impossible because adding the context-aware protocol does not change the core functionality of the permission sequence protocol. Having an ESO capability does not give the attack any advantage in violating the permission sequence. However, we discuss the following new attack scenario.

{\bf{\textit{Client impersonates as RS. }}} In our system design, the RS contacts the ESOs to check the environmental conditions during policy enforcement. In fact, the client should not have the right to obtain the state of the context via ESOs. However, the malicious client may query the context state from the ESOs by directly sending the ESO capability.  To prevent such attack, we require that an ESO must authenticate the RS and verify that the RS is authorized to access by checking the ESO capability. An RS can authenticate to an ESO  by proving the possession  of the private key (e.g., Signing $\mathit{T_{Con}}$ and sending the capability with its signature to the ESO).

Capabilities, server responses should be kept confidential in transit (using TLS/DTLS), at least between the RS and ESO. Since ESO responds sensitive data about the user's activity.

\section{Limitations}\label{limit}
\noindent\paragraph{\textbf{Storing additional state.}} The protocol requires the resource servers to maintain an additional state for each client. The resource servers also need sufficient computational power to verify the signatures on the tokens and digitally sign the new ones. In IoT applications, this may be a problem if a resource-constrained resource server must serve many users. This is likely not a problem in Web applications where servers have access to storage and computing to serve many users. Compared to HCAP \cite{tandon2018hcap} considers enforcement of general history-based policies, the information that is stored on the resource server is significantly reduced. The resource server only needs to maintain an internal counter.

\noindent\paragraph{\textbf{Reusability of tokens.}} The ubiquitous use of tokens is due to its flexibility and ease of adoption. For example, tokens can be reused to access protected resources. As discussed in Section \ref{intro}, reusing tokens, however, can pose a significant security risk.  Our proposal can be used to control the order and the number of token use while maintaining the system's security, and the price to pay is that the AS needs to generate a new master token for each session.  To optimize the design in reducing the number of interactions with the AS,  all descendant tokens in the same session are derived from the master token directly without contacting the AS again. HCAP \cite{tandon2018hcap} already bounds tokens to sessions. Therefore, our proposal is not making the efficiency worse when controlling the order and the number of token use.

\section{Implementation}\label{imple}
We implement our proposed system based on the OAuth client credential grant with proof of possession tokens. OAuth has two token structures,  bearer tokens and proof of possession (PoP) tokens. Unlike bearer tokens, PoP tokens are tied to a specific subject, and the requesting party has to prove the possession of a secret key only known to this subject, hence aligns with capabilities in our system. Our theoretical system is not restricted to the use of any specific access control model. To demonstrate the practical use of our capability system, we implement ABAC \cite{hu2013guide} as the authorization mechanism in AS. ABAC shows supremacy on scalability because policies are defined for subject-object attribute combinations. On the technical side, each server is developed using Node with Express framework and MongoDB database.

 We discuss the implemented components in Section \ref{AS} and  \ref{RS}. We then present a use case that shows how the two features of the proposed scheme are useful in distributed financial systems (Section \ref{WholeExample}).

\subsection{Authorization Server}
\label{AS}The AS provides the authorization and token generation service. We implement ABAC model \cite{hu2013guide} for authorization and JSON web token \cite{jsonwebtoken} as the token format.

{\em{\bf{Policy Language Model}}}
Based on the theoretical design in Sections \ref{permission} and \ref{Context}. We define policies as the following model:\\
\noindent$\langle \mathit{Policy} \rangle ::=\langle \mathit{Rule}\rangle | \langle \mathit{Rule} \rangle \langle \mathit{Policy}\rangle$\\
$\langle Rule \rangle::= \langle \mathit{subjectAttributes} \rangle \langle \mathit{objectAttributes} \rangle \langle \mathit{authorization} \rangle\\
\hspace*{1.35cm}\langle \mathit{actionAttributes} \rangle\langle \mathit{environmentContext} \rangle\langle \mathit{Default} \rangle$\\
The $\langle \mathit{Policy} \rangle$ consists of a set of $\langle \mathit{Rule}\rangle$s. A $\langle \mathit{Rule}\rangle$ must conform to the following form:
\begin{itemize}[label={--}]
	\item $\langle \mathit{subjectAttributes} \rangle$: Attributes of the clients.\\
	$\langle \mathit{objectAttributes} \rangle$: Attributes of the objects. The objects are the resources or services protected by the resource servers.
	\item $\langle \mathit{authorization} \rangle$: The result of policy evaluation. Either ``permit''  or ``deny'' .
	\item $\langle \mathit{actionAttributes} \rangle$ specifies permissions and their scope. 
	\item $\langle \mathit{environmentContext} \rangle$ specifies any external context. 
	 Note that we separate the environment context from the permission for easier demonstration.  Evaluation of context is performed by the resource server in conjunction with the enforcement of an authorization decision.
	\item $\langle \mathit{Default} \rangle$: Indicating the default decision if the attributes of the requester or attributes of the object do not match with the attributes in $\langle \mathit{Rule} \rangle$.
\end{itemize}

{\em{\bf{Policy Implementation}}} The AS  support any policy content as long as it conforms to the policy language model. 
It stores the policies as a collection of rules in MongoDB. We create Policy Schema in MongoDB based on the policy language model. The rules are implemented as JSON objects. \textit{ We show a policy rule example below and its implementation in Figure \ref{policy1}.}  

\textit{Example.} Alice uses Application B that requires a paid membership.
 Application B offers Alice the option to pay her membership monthly using her credit card. Alice authorizes her credit card company to pay the application fee under the following conditions.   Application B can make once a month \$10 charge to Alice’s account, under the condition that Alice has been using Application  B for the past two months. Thus a payment request will be rejected in the following cases, \ding{202} Application B is requesting an amount different from \$10. \ding{203} Application B is charging \$10 to Alice’s account for the second time in the same month. \ding{204} Alice has stopped using Application B, but she has not canceled her subscription. This last case will be detected by monitoring access to the application.

\begin{figure}[t]
\begin{lstlisting}[language=json]
{  
    "type":"ABAC policy",
    "name":"ApplicationServiceCharge",  
    "application":"Payment",
    "rules":{
        "subjectAttribute":{
            "ApplicationID":["B"]
        },
        "objectAttribute":{
            "resourceType":["balance"],
            "resourceID": "Alice"
        },
        "authorization":"permit",
        "actionAttribute":{
            "actions":["charge"],
            "amount": "$10",
            "frequency": "monthly"
        }, 
        "environmentcontext":["used_within_two_months"],
        "Default":{
            "authorization":"deny"
        }
    }
}
\end{lstlisting}
\caption{Allowing one-time charge authorization with \$10 when Alice has been using Application  B for the past two months.}\label{policy1}
\end{figure}

\paragraph{\bf{Authorizaton Evaluation}} The client is authenticated by signing the client claim token. Then, the AS decodes the client claim token, gets the client identity, and queries the subject attribute database with this identity. With the found subject attributes, the AS evaluates it against all the available policies. More specifically, the AS checks \ding{202} if the requester has all the attributes required in $ \langle \mathit{subjectAttributes} \rangle$ \ding{203} if the requested object are allowed by the $ \langle \mathit{objectAttributes} \rangle$  \ding{204} if the requested scopes/permissions are allowed by the $ \langle \mathit{actionAttributes} \rangle$. The AS adopts the permit-override strategy. If there exists at least one permit, the AS will grant access and generate a master token. To generate the master token, the AS pulls out the relevant information in the policy and assembles them in the token. 

If any environmental context is required by the policy, the AS will need to create an ESO token which is used to query a responsible ESO server. When an ESO server is introduced to the AS, AS will record its description and URL in the known ESO ``list'' . With this ``list'' , the AS can find the ESO server based on   the context from policy $\langle \mathit{environmentContext} \rangle$. For example, the ``used\_within\_two\_months''  in Figure \ref{policy1} maps to the ESO that checks if the user has used the app within the last two months.

\subsection{Resource Server and ESO Server}\label{RS}
Object attributes are bound to their objects through referencing, by embedding them within the object. In our case, the resources are objects, and they are stored together with their attributes. The policy enforcement point is implemented in the RS. Upon receiving the tokens from the client, the RS verifies the signatures of each token, contents of each token, and checks if two tokens are bound. If context validation is required, the RS sends an HTTPS post request to fetch the internal state of the ESO. 

\begin{figure}[H]
\begin{lstlisting}[language=json]
{  
    "client_id":"B",
    "issuer":"ApplicationB",  
    "application":"Payment",
    "objectAttribute":{
        "resourceType":["balance"],
        "resourceID": "Alice"
    },
    "structured_scope":{
        "actions":["charge"],
        "amount": "$10"
    }
}\end{lstlisting}
\caption{Example of decoded client claim token}\label{client-assertion}
\end{figure}
Once all these steps are successful, the RS may return the resources, or provide services to the client. We wrap all the token validations and context checking duties into an Express middleware. Programmers can easily invoke this independent module when writing any customized API for the RS.

We do not implement any underlining protocols for tracking environment context as it is out of the scope of this work. After successful validation of the ESO token and the RS signature, the ESO server sends a response \{``Contex'' :``True'' \} or \{``Contex'' :``False'' \} to the RS.

\subsection{Use Case}
\label{WholeExample}

\paragraph{\bf{Client authorization request}}
App B requests authorization from Alice's bank to charge her account \$10 per month. App B sends an HTTPS request with the following keys in the header to the AS that the bank maintains. Figure \ref {client-assertion} shows the decoded client claim token in ``\textbf{client-assertion}''.

\begin{tcolorbox}[colback=gray!5!white,colframe=gray!75!black]
``\textbf{grant-type}''  :``client\_credentials'',\\ ``\textbf{client-assertion-type}'' : ``jwt-bearer'',\\ ``\textbf{client-assertion}'' :``eyJhbGciOiJFUzI1NiIsInR5cCI6IkpX\\VCJ9...jgP0WPftkxaYg5LjVCS4Q2Dp6hQ''.  
\end{tcolorbox}

\paragraph{\bf{Authorization Server Response}}
Figure \ref{subjectattribute} shows the attributes of App B. The authorization server evaluates the client request
against the policy and uses the rule that permits one-time
charge of \$10 if the user has used the App in the last two months.
In this example, the AS issues both the master token and the ESO token. Figure \ref{master} shows the decoded master token, and Figure \ref{esotoken} shows the decoded ESO token. 

The master token contains the information of the token expiration time, subject, audience (RS URL), issuer (AS endpoint), state, actionAttributes, and the required environment context. 

ESO token contains the hash of the master token (we use SHA256 implementation from CryptoJS library \cite{mott2015crypto}), expiration date, subject, audience (ESO server URL), issuer (AS endpoint), and the required environment context.

\begin{figure}[t]
\setlength{\belowcaptionskip}{-3.5pt}
\begin{lstlisting}[language=json]
{  
    "subject_id":"B",
    "application":"Payment",
    "subjectAttribute":{
        "ApplicationID":["B"]
    },
    "name":"ApplicationB"
}\end{lstlisting}
\caption{Example of client attributes }\label{subjectattribute}
\end{figure}

\begin{figure}[t]
\setlength{\belowcaptionskip}{-3.5pt}
\begin{lstlisting}[language=json]
{   
    "expireIn":"1 day",
    "subject_id":"B",
    "audience":"https://localhost:4990/Alice/balance",
    "issuer":"https://localhost:5000/authorization",
    "state":"0",
    "actionAttribute":{
        "permission_sequences":["charge"],
        "amount": "$10",
        "frequency":"monthly"
    }, 
    "environmentcontext":["used_within_two_months"],
    "iat":1567468693
}\end{lstlisting}
\caption{Example of decoded master token }\label{master}
\end{figure}

\begin{figure}[H]
\setlength{\belowcaptionskip}{-3.5pt}
\begin{lstlisting}[language=json]
{   
    "expireIn":"1 day",
    "hashAT":{
        "words":[
            1904756807,
            -1499235065,
            -860331953,
            -1557528208,
            -355723369,
            -1355021346,
            -70944964,
            -653925533
        ],
        sigBytes":32
    },
    "subject":"https://localhost:4990/Alice/balance",
    "audience":"https://localhost:4995/used_within_two_months",
    "issuer":"https://localhost:5000/authorization",
    "action":["read"],
    "userID": "Alice",
    "environmentContext":["used_within_two_months"],
    "iat":1567468693
}\end{lstlisting}
\caption{Example of decoded ESO token }\label{esotoken}
\end{figure}

\paragraph{\bf{Resource Request}}  The client App B sends an HTTPS request with the following keys in the HTTP headers to the RS endpoint that manages customers' accounts,
\begin{tcolorbox}[colback=gray!5!white,colframe=gray!75!black]
`\textbf{`x-oauth-token'' } : ``$eyJhbGciOiJFUzI1NiIsInR5c...Uoly\\kHKeJUHcGho2A$'',\\ ``\textbf{x-eso-}\textbf{token}''  :``$eyJhbGciOiJFUzI1NiIsInR5c...XBCsXE\\jgq8XWuLpXg$''.
\end{tcolorbox}
After all the checks have passed based on Algorithm \ref{alg1} , RS queries the ESO that tracks Alice's login history of App B in the past two months . If the ESO server respond \{``Context'' :``True'' \}, the RS will add \$10 charge to her account. Finally, the RS will inform the AS about the transaction. The AS will not issue any new token until the next billing cycle. 

\section{Performance Evaluation}\label{eva}
We evaluate the implemented system for two token signature algorithms, RSA signature, and ECDSA signature. ECDSA signature algorithm has a smaller key size compared to the RSA signature algorithm for the same level of security. With the faster signing process and smaller key storage, the ECDSA signature can be used for constrained devices, such as IoT devices. The RSA operations and elliptic curve operations related to the token generation and verification are implemented using the jsonwebtoken library\cite{jwt}. The key length in RSA signature is 3072 bits, and the curve in ECDSA is P-256. Both of them provide 128 bits security\cite{suarez2018practical}. The hash functions are SHA-256.

\begin{figure*}[t]
	\centering
	\subfloat[Average response time (in milliseconds) for authorization request through our system with 95\% confidence interval. \label{ourSystemAS}]{\includegraphics[width=.4\textwidth]{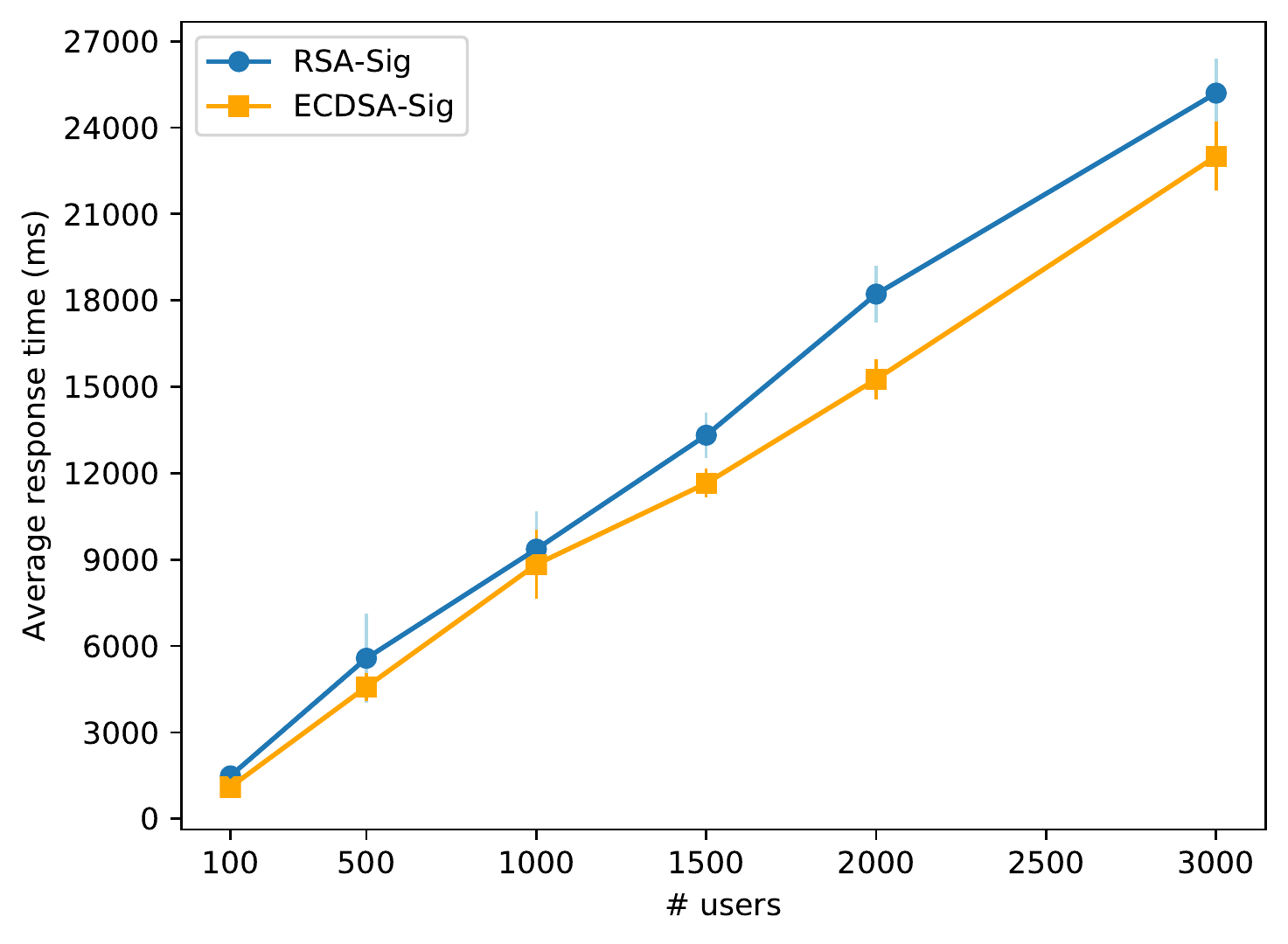}}
	\hspace{3mm}
	\subfloat[Average response time (in milliseconds) for the resource request through our system with 95\% confidence interval. \label{ourSystemRS}]{\includegraphics[width=.4\textwidth]{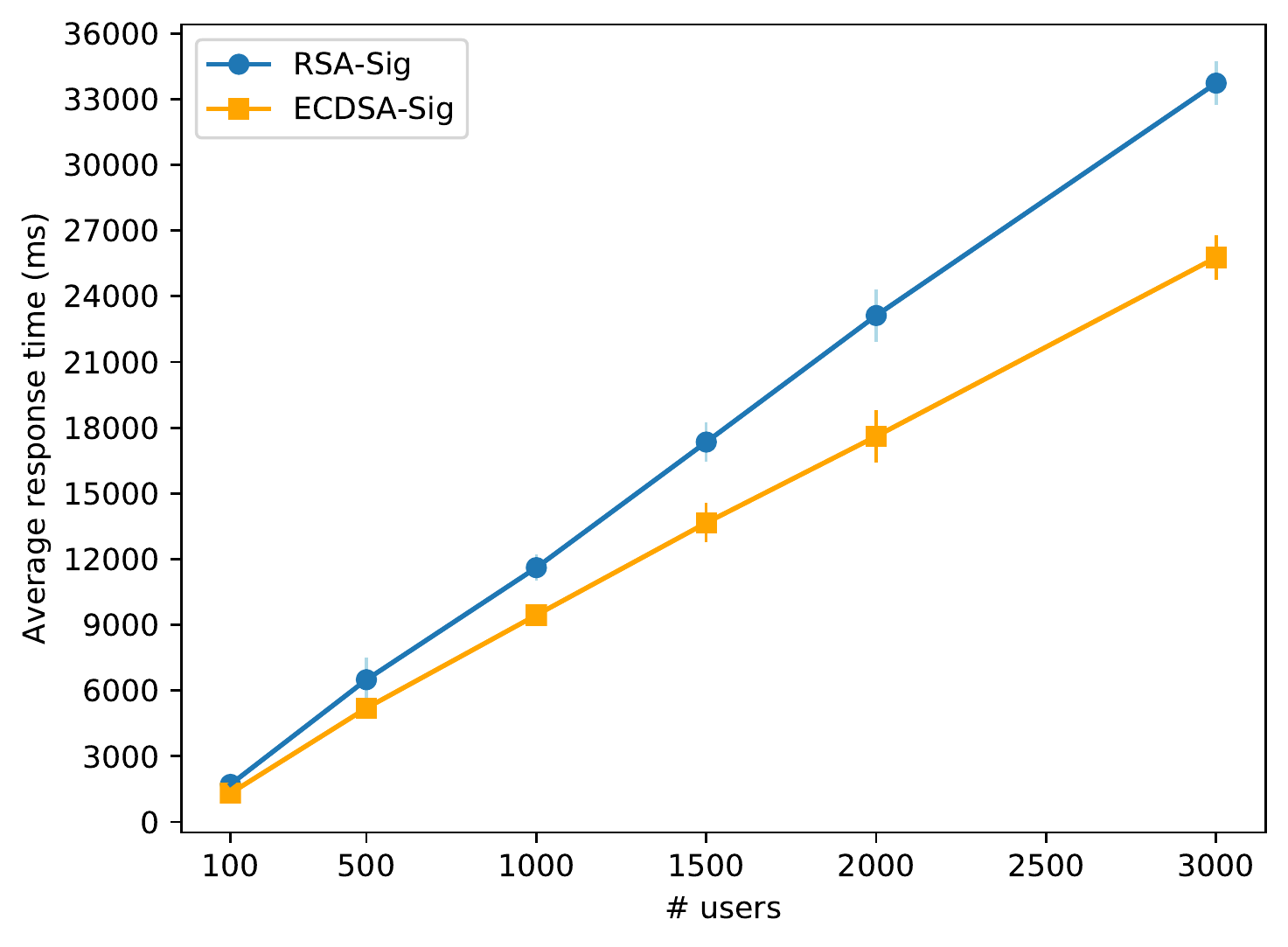}}
	
	\subfloat[Multiplicative overhead of the authorization request:  average response time in our system compared to OAuth.\label{overhead1}]{\includegraphics[width=.4\textwidth]{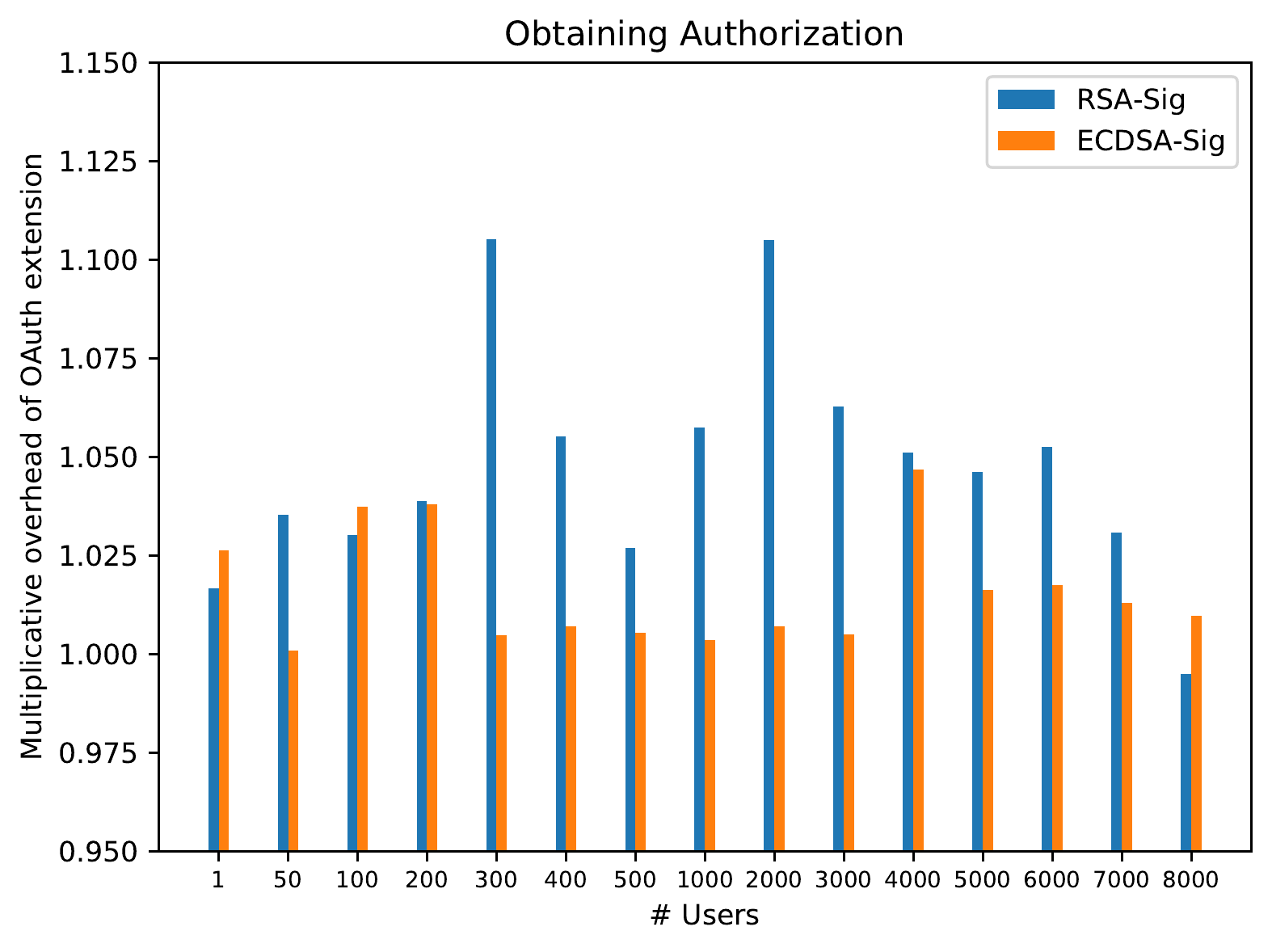}}
     \hspace{3mm}
	\subfloat[Multiplicative overhead of the resource request:  average response time in our system compared to OAuth. \label{overhead2}]{\includegraphics[width=.4\textwidth]{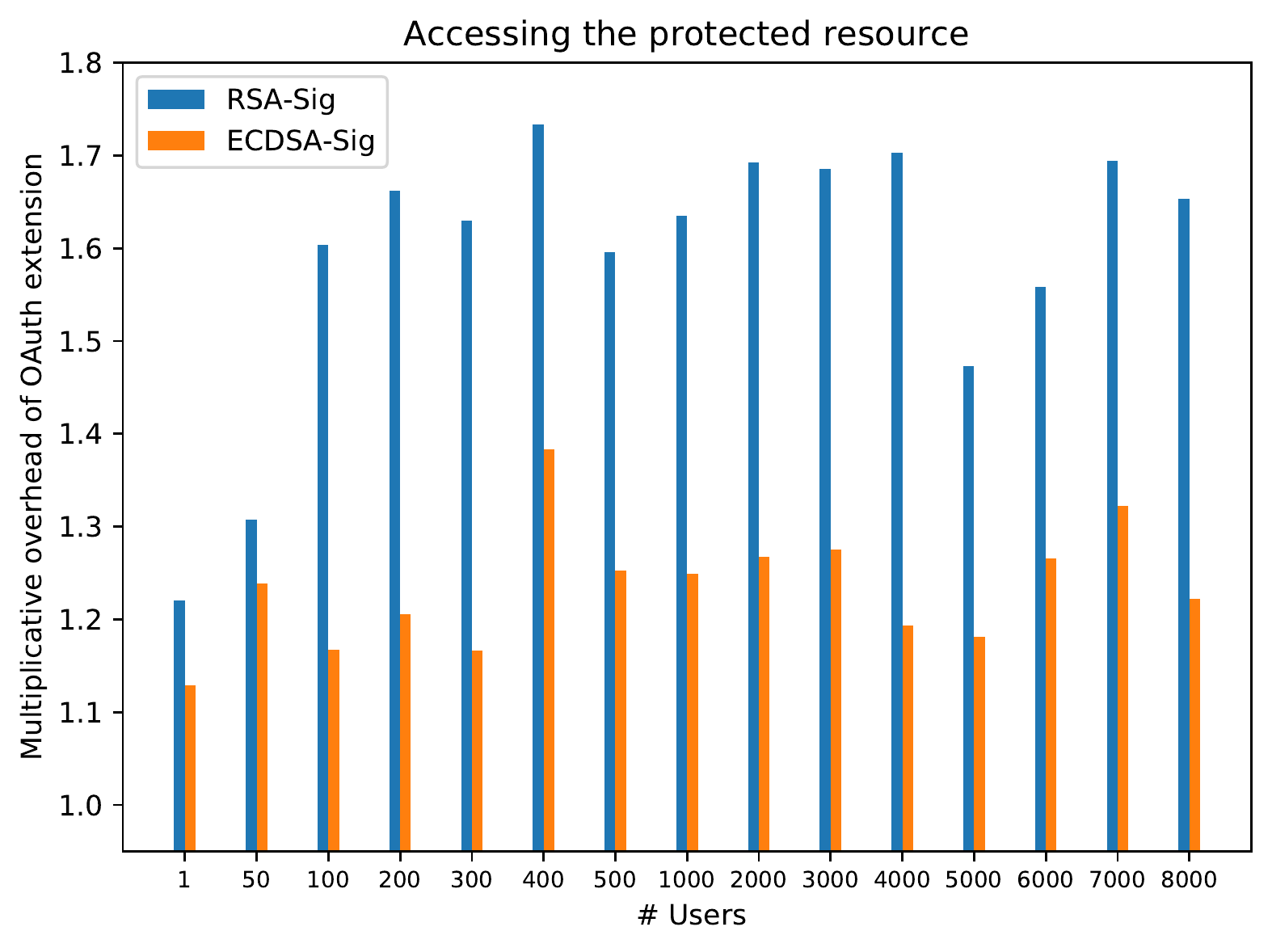}}\
	
	\caption{Performance of two requests in our system and their overhead compared with OAuth.}
	\label{fig:corr}
\end{figure*}

To evaluate the performance, we implement various settings described below. The experiments are run on a windows computer with  4 Core(s) 8 Logical Processor(s) 3.60GHz    Intel(R) Core(TM) i7-7700 CPU and 8GB of RAM. We use Apache JMeter as the client side to send https requests to the AS and the RS. Both servers and the ESO server run on different ports of the same computer. The client and the servers reside on the same computer with a negligible round-trip latency (0.1-0.2ms). All communications are TLS protected. The TLS version is TLSv1.2, and the cipher suite is TLS\_ECDHE\_RSA\_WITH\_AES\_128\_GCM\_SHA256. We use X.509 certificates for TLS authentication.

The goal of the experiment is to answer the following questions.  \ding{202} How does the number of simultaneous requests impact the average response time? \ding{203} How long does it take to receive a response from the
server? \ding{204} Compared to OAuth 2.0, how much enabling 
context and permission sequence constraints slows the authorization process and access? 

We address these questions in terms of two requests, the authorization request to AS, and the resource request to RS. For each type of requests, we first present the average response time for a varying number of simultaneous requests. Then we examine the time breakdown for one request. Finally, we compare our system with OAuth 2.0 and show the overhead based on the average response time.

\subsection { Authorization Requests}\label{8.1}

\paragraph{ \textbf{The average response time of simultaneous requests}} We report the average response time for up to 3000 simultaneous authorization requests in Figure \ref{ourSystemAS}. The reported result is an average of 5 experiments. Figure \ref{ourSystemAS} also shows the 95\% confidence interval calculated from 5 experiments. We use the request example in Section \ref{WholeExample} for all the requests and evaluates the request against 10 ABAC policies. 

To ensure accuracy, we use the Apache JMeter to automate our test. JMeter creates a new thread for each user created.  Each user is independent of another user. We configure $n$ users, and each user fires off one request. These $n$ requests are sent in series within one second.

As shown in Figure \ref{ourSystemAS}, the average response times in the ECDSA-P256 setting are shorter, compared with RSA SHA-256 setting. This is because creating RSA signature takes significantly more time.  

\begin{table}[h]
	\centering
	\caption{Runtime breakdown in an authorization request (in milliseconds)}
	\label{ourSystemASbreakdown}
	\begin{tabular}{ccc}
		\toprule
		\textbf{Operations} & \textbf{RSA-Sig} & \textbf{ECDSA-Sig}\\
		&\textbf{(time in ms)}&\textbf{(time in ms)}\\
		\midrule
		TLS handshake &	12 &	13.2\\ 
		Client Authentication &	0.7    &	0.942\\
		Check Permissions &   73.381 & 75.56\\
		master token & 4.318 & 0.508\\
		ESO token & 4.266 & 0.486 \\
		Estimated Total Time & 94.665 &	90.696\\
		\bottomrule
	\end{tabular}
\end{table}

\paragraph{\textbf{Time breakdown  of a single request}}\label{8.2}

Table \ref{ourSystemASbreakdown} gives the time break down for the case of  RSA-Sig and the ECDSA-Sig. We also report the time duration of the TLS handshake. All numbers are the average of over 1000 token requests. 

In our experiments, most of the processing time is spent on checking permissions, which includes the time for retrieving context information to be able to take the access decision. The time of checking permission is similar in the RSA-sig setting and ECDSA-sig setting. This aligns with our implementation, as checking permissions does not include signature operations.  The computation time of creating the master token and ESO token is similar for the same signature algorithm. On the other hand, the computation time of creating a token with RSA-Sig is significantly more than creating a token with ECDSA-Sig.

\paragraph{\textbf{ Comparison with OAuth 2.0 on authorization requests}}
We implement the OAuth client credential grant for comparison. Comparing the average response time of authorization request in our system and OAuth 2.0 presents the extra cost of enabling environment context on the AS side. We show the multiplicative overhead of the average round trip time of the authorization request in Figure \ref{overhead1}. Different token signatures represent in different colors. The set of bars represents the overhead of our system compared to OAuth for various number of requests.  We evaluate the overhead up to 8000 requests. 

In the AS, since the most-time consuming component is checking permissions, the overhead of our system compared with OAuth is fairly small. As shown in Figure \ref{overhead1}, the overhead is less than 5\% in the ECDSA-Sig setting and 12\% in the RSA-Sig setting.

\subsection{Resource Requests }\label{8.4}
\paragraph{ \textbf{The average response time of simultaneous requests}}  We report the average round-trip return time for up to 3000 simultaneous resource requests in Figure \ref{ourSystemRS}.  The reported result is an average of 5 experiments. Figure \ref{ourSystemRS} also shows the 95\% confidence interval calculated from 5 experiments. We use the request example in Section \ref{WholeExample} for all the requests and evaluates the request against 60 user profiles.

The response time grows with the request $n$. The average response times in the ECDSA-Sig setting is shorter compared with the RSA-Sig setting. This is because when the RS checks the context with the ESO server, it signs the ESO toke and sends the token with the signature to the ESO server. As we know, creating RSA signature takes significantly more time. This signature is necessary as it prevents the client impersonation attack in Section \ref{Context}.

\paragraph{\textbf{Time breakdown  of a single request}}\label{8.5}
Similarly,  we report the time break down for one resource request. As shown in Table \ref{ourSystemRSbreakdown}. The time of checking the hash, token validation, resource query are similar for different token signatures, since those operations do not include any signature operations. 

For the same signature algorithm, the computation time for verifying the signature of the master token and the ESO token does not vary. On the other hand, the computation time for verifying the RSA signature is less than that of the ECDSA signature.

\begin{table}[h]
	\centering
		\caption{Runtime breakdown in a resource request (in milliseconds)}\label{ourSystemRSbreakdown}
	\begin{tabular}{ccc}
		\toprule
		\textbf{Operations} & \textbf{RSA-Sig} & \textbf{ECDSA-Sig}  \\
		&\textbf{(time in ms)} &\textbf{(time in ms)}\\
		\midrule
		
		TLS handshake &	14.2 &	13.5   \\ 
		master token signature verify &	0.36    &0.49  \\
		ESO token signature verify	&   0.34	&0.5\\
		Check Hash	&   0.401 &	0.329	 \\
		Token Validation	&   0.006	&0.004 \\
		Context checking &20.9&15.0 \\
		Resource query & 66.679 & 66.458\\
		Estimated Total Time & 102.886 & 96.281\\	
		\bottomrule
	\end{tabular}

\end{table}

\paragraph{\textbf{Comparison with OAuth 2.0 on resource requests}}
 We removed the codes of ESO token signature validation, hash check and context checking. The modified RS checks the signature and validates the master token. Then it invokes the resource query function.
Comparing the average response time of resource request in our system and OAuth 2.0 presents the extra cost of enabling context on the RS side. 
The multiplicative overhead of our system is shown in Figure \ref{overhead2}. The overhead for RSA-Sig and ECDSA-Sig remains a small constant.

\section{Related Work}\label{work}

Our work is built upon a wealth of prior research in the field of distributed authorization. In this section, we discuss the most related lines of work and present the key ideas involved in the design of our capability system.

Capability-based mechanisms and credential-based mechanisms are the two main approaches to solving the distributed authorization problem \cite{birgisson2014macaroons, clarke2001certificate, gong1989secure, gusmeroli2013capability, hardt2012oauth, 6915840, maler2016user, sciancalepore2017oauth, ACE}. The key concept in capability-based based authorization is the capability, which is defined in \cite{dennis1966programming} as ``token, ticket, or key that gives the possessor permission to access an entity or object in a computer system''. The subject who wishes to access the resource custodian sends a capability together with the request. 
The notable examples of such system  are OAuth 2.0 \cite{hardt2012oauth}, UMA\cite{maler2016user} and ICAP\cite{gong1989secure}. In credential-based authorization, a target service determines a subject's authorization to invoke permission by examining assertions encoded in verifiable digital credentials, which are usually issued by different authorities. The notable examples are SPKI/SDSI \cite{clarke2001certificate} and Macaroons \cite{birgisson2014macaroons}. This work contributes to the work of capability-based authorization.

For context constraints, several distributed capability system supports enforcing local context, such as \cite{gusmeroli2013capability, hernandez2016dcapbac}. Popular authorization protocols like  OAuth 2.0 and UMA  already support a common library of contexts in their implementations, such as specifying the expiration time and the client's identity. To support complex conditions, the adoption of $\mathit{structured\_scope}$ assertion in OAuth token was proposed in \cite{ mohammad, Tor}. 
$\mathit{Structured\_scope}$ is a JSON object which specifies the permissions with the conditions rather than a plain string of permissions.
However, this only works for the local conditions, not for  the environmental context. 

In \cite{schuster2018situational}, the authors proposed a way to enforce environment context in IoT. They introduced the Environmental Situational Oracle (ESO), which tracks the context and is entirely external to any entity. The ESO exposes a simple interface that can be used to determine whether the situation is true or not. They also proposed a two-round OAuth flow, which allows the RS to verify the context with the ESO.
This paper uses the ESO but reduces the protocol to only
one round of OAuth 2.0 by using two capabilities with different delegated permissions. Our proposed protocol requires one less round of OAuth flows but keeps access to the ESO protected. External context service providers have been proposed to aid context-aware resource sharing in Cloud environment \cite{kaluvuri2015safax} and Intelligent Transport Systems \cite{ravidas2019authorization}. Compared to these two works, we add one extra layer of protection to the context providers by requiring tokens upon access, as they usually track sensitive information about users. 

On the other hand, in the credential-based authorization.  Birgisson et al. proposed Macaroons \cite{birgisson2014macaroons}, which are credentials with contextual confinement. The target service issues Macaroons to the client, and then the client uses Macaroons and other necessary claims to obtain access from the target service. The permissions and their contextual confinement are embedded in the Macaroons. 
Compared to \cite{birgisson2014macaroons}, our capability system is compliant with the distribute setting (entities, message flows) considered by OAuth 2.0 and UMA. Hence, it is easier to extend these authorization protocols with our system to have granular policies. 

The authorization frameworks mentioned above offer no control over permission sequences constraints. In capability-based systems such as ICAP and OAuth, capabilities may carry a list of granted permissions. These permissions do not have any usage constraints.
Recent work in \cite{tandon2018hcap} presents a capability system (HCAP) that supports history-based access control. Compared to HCAP, our work has several advantages when enforcing a finite permission sequence. First, any communication between resource servers is not required. Second, HCAP uses the timestamp to invalidate the old capability, which requires a synchronized clock. We do not have this assumption in our capability system. Third, HCAP capabilities carry part of the request security automaton (a security automaton specifies the enforceable security policy), and the RS is required to simulate the state transition in the security automaton. Our capabilities carry the finite permission sequence, and the RS simply needs to maintain an internal counter. The lightweight capabilities and fast capability verification pave the way for faster deployment of our capability-based system for ecosystems such as IoT. Finally, the capability in HCAP  includes a Message Authentication Code using the shared secret. This security tag is only verifiable by the target resource server. In this work, the signatures in capabilities are public-key-based thus are public-verifiable. It is worth noting that HCAP does not support ``context''  of access.

\section{Conclusion}\label{Con}

The effectiveness of applying capability-based systems to access control in many domains depends on its ability to enforce complex policies, including orderings among permissions and environmental situations. However, efficient enforcement of these conditions in a distributed capability system is challenging since policy decision and enforcement are carried out in different entities. We motivate this research to provide a capability-based system with the fine-grained delegation of authority and efficient enforcement of conditional constraints. 

We described a capability-based system, which supports enforcement of permission sequence and context constraints. Capabilities in our model allow the resource owner to have control over orderings among permissions and specify any external conditions in the policies. Cryptographic means in the capability system provide advanced security features and efficient capability revocation. We formally proved the safety property of the proposed system. We integrated our system with OAuth 2.0 and demonstrated that the performance of our system is competitive. Future work will focus on enforcing the other history-based policies using minimum state. Furthermore, we will consider an honest but curious RS and ensure that the RS can not passively/actively learn more information about the user and their surrounding environment.

\section{Acknowledgements}
This research is in part supported by Natural Sciences and Engineering Research Council of Canada and Telus Communications, under Industrial Research Chair Grant scheme.

\bibliographystyle{abbrv}

\bibliography{BibFile} 

\end{document}